\documentclass{kapproc} 

\usepackage{t1enc}

\usepackage{procps} 

\usepackage[dvips]{graphicx}

\upperandlowercase

\kluwerbib 

\begin{document}

\articletitle{Ionising stellar populations in\\
 circumnuclear star-forming regions}


\author{Enrique P\'erez-Montero}
\affil{Dpto. de F\'\i sica Te\'orica. Universidad Aut\'onoma de Madrid. Spain.}
\email{enrique.perez@uam.es}

\author{\'Angeles I. D\'\i az}
\affil{Dpto. de F\'\i sica Te\'orica. Universidad Aut\'onoma de Madrid. Spain.}
\email{angeles.diaz@uam.es}

\author{Marcelo Castellanos}
\affil{Laboratoire d'Astrophysique de Toulouse-Tarbes. Observatoire Midi-Pyr\'en\'ees. France.}
\email{castella@ast.obs-mip.fr}

\section{Circumnuclear Star Forming Regions}
Circumnuclear Star Forming Regions (CNSFR), also called hotspots, can be found 
in the inner parts of disc galaxies (r < 1 kpc) where intense processes of star 
formation are taking place, associated to rings or pseudo-rings. They appear to be 
more compact and with a more pronounced peak in the surface brightness distribution 
than other types of HII regions. Regarding the properties of their stellar ionising 
populations, \'Alvarez-\'Alvarez, D\'\i az \& Castellanos (2003) needed a composite population 
with both a just-formed burst and an older one in the WR-phase in order to explain 
some emission-line ratios, Wolf-Rayet features and blue colors. There is a coincidence 
in this point with some recent results for HII galaxies (P\'erez-Montero \& D\'\i az, 2004). 
Some of the properties of CNSFRs have been studied, analyzing a sample of them through 
spectrophotometrical observations, using diagnostic emission-line ratios involving 
oxygen and sulphur. The deduced properties have been compared with those of other 
families of objects and results from photo-ionisation models.

\section{The sample of studied objects}
We have selected a sample of emission line-objects from the literature with [OII], 
[OIII], [SII] and [SIII] emission lines, thus obtaining different empirical calibrators 
of the functional parameters, for different families of HII regions, including Giant 
extragalactic HII regions and HII galaxies. With the aim of comparing these properties 
with those of CNSFRs, we have compiled data from P\'erez-Olea (1996) with observations 
of CNSFRs in NGC2903, NGC3351 and NGC3504, that include these lines as well. Data for 
other CNSFRs (M51; D\'\i az et al. 1991, NGC 3310; Pastoriza et al. 1993 and NGC 7714; 
Gonz\'alez-Delgado et al. 1995) have been added too.

\section{Functional parameters}
Photoionisation equilibrium in HII regions is governed by three main parameters 
so-called functional: the metallicity, the ionisation parameter (i.e. the quotient 
of ionising photons to the density of particles) and the hardness of the incident 
continuum. In previous works it has been shown that the estimation of these parameters 
through the sulphur emission lines provide values of them  with a high level of confidence.
    
\begin{figure}[h]
  \begin{center}
    \vspace{0.8cm}
    \includegraphics[width=7cm]{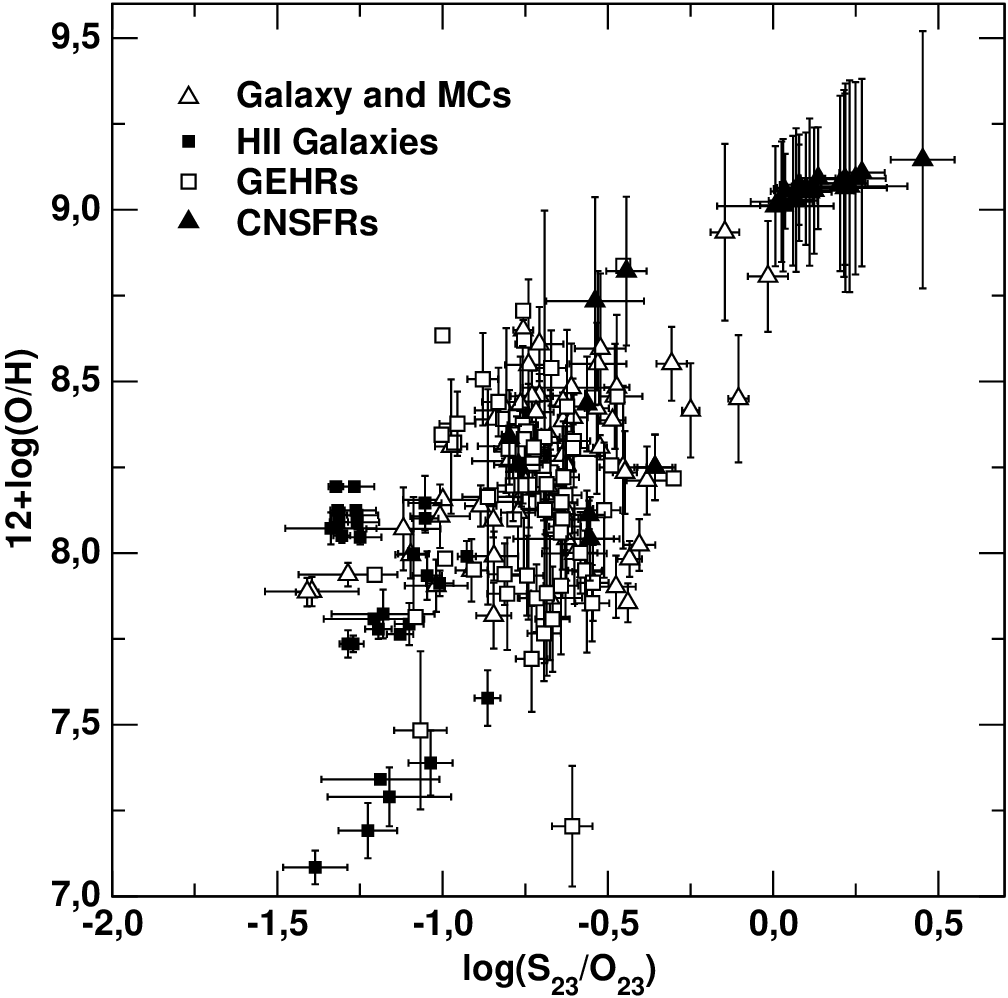}
  \end{center}
\caption{Relation between the metallicity, in terms of 12+log(O/H), and the S23/O23 parameter
for different families of HII regions, including CNSFRS.}
\end{figure}

\subsection{Metallicity}
Many CNSFRs are objects with very large metallicites ($Z \geq Z_{\odot}$), and it is no 
possible to ascertain their oxygen abundances through the direct method and, therefore, 
it is compulsory to use empirical parameters. For this sample we have used the parameter 
S23/O23 (D\'\i az \& P\'erez-Montero, 2000), that leads to values of oxygen abundance at high 
metallicities with little uncertainty. For the sample of CNSFRS from P\'erez-Olea, 
, upper right corner in figure 1, we obtain values 
between 1.09 and 2.88 $Z_{\odot}$. On the other hand, the CNSFRS with low metallicities, for 
which it is possible to apply the direct method, present abundances between 0.22 and 
0.55 $Z_{\odot}$. 

\begin{figure}[h]
  \begin{center}
    \vspace{0.8cm}
    \includegraphics[width=7cm]{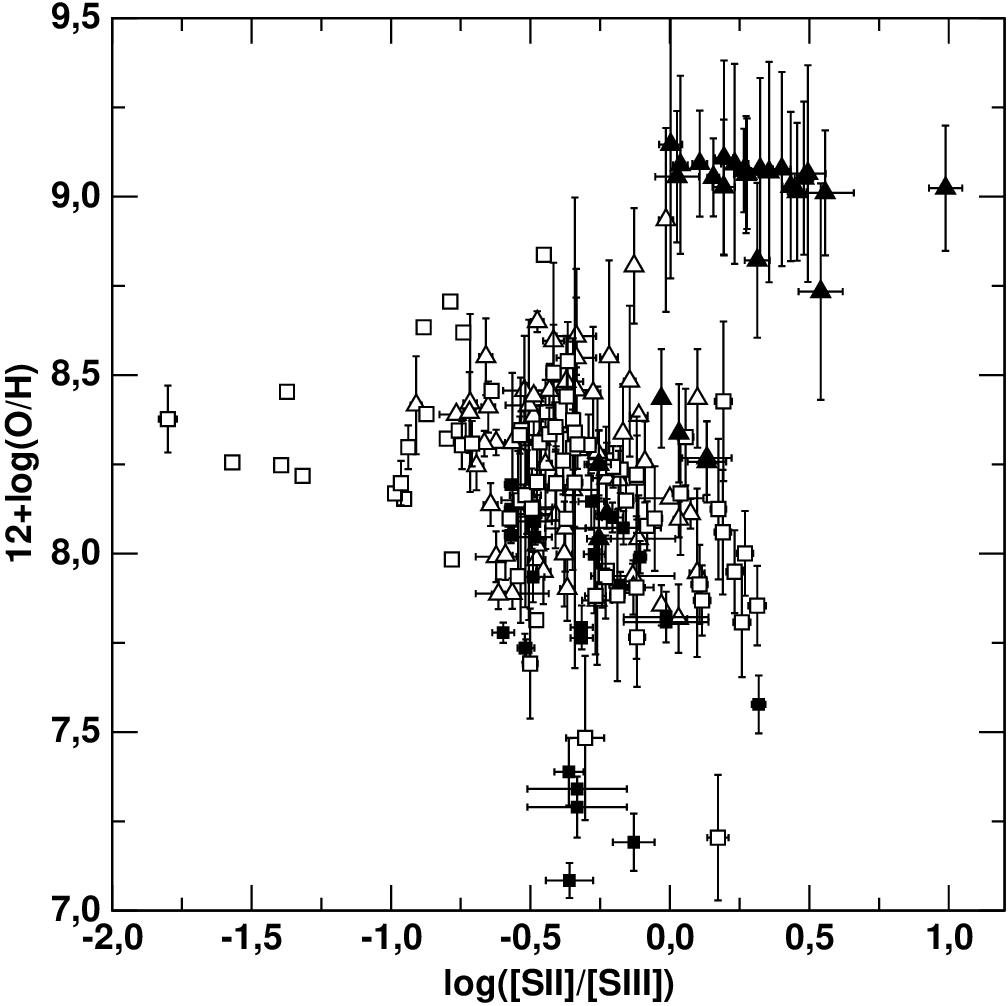}
  \end{center}
\caption{Relation between [SII]/[SIII] and the metallicity. For CNSFRS, this last obtained from the
S23/O23 parameter.}
\end{figure}

\subsection{Ionisation parameter}
It gives an idea of the excitation degree inside the nebula. Although, the [OII]/[OIII] 
ratio has been used frequently to parametrise it, we have used [SII]/[SIII] 
(D\'\i az et al., 1991), because this ratio has no dependence on effective temperature. 
This parameter, as we can see in figure 2, does not present any correlation 
with metallicity. CNSFRs, contrary to HII galaxies, tend to present large values of 
[SII]/[SIII], what implies low degrees of excitation.

\subsection{Effective temperature}
The temperature of the field of radiation can be deduced from the parameter $\eta$ ' 
(V\'\i lchez \& Pagel, 1988), that is defined as a function of the bright emission
lines of oxygen and sulphur:
\[ \log \eta' = \frac{\log([OII])/\log([OIII])}{\log([SII])/\log([SIII])} \]
As we can see in figure 3, there seems to be a sequence defined by this 
temperature and the metallicity, represented by S23/O23 in this diagram, in such 
a way that, for instance, low metallicities correspond to higher effective equivalent 
temperatures. This is to be expected from stellar atmospheres theory. Nevertheless, 
CNSFRs are an exception because, independently of their oxygen abundance they present 
always very high effective temperatures.

\begin{figure}[h]
  \begin{center}
    \vspace{0.8cm}
    \includegraphics[width=7cm]{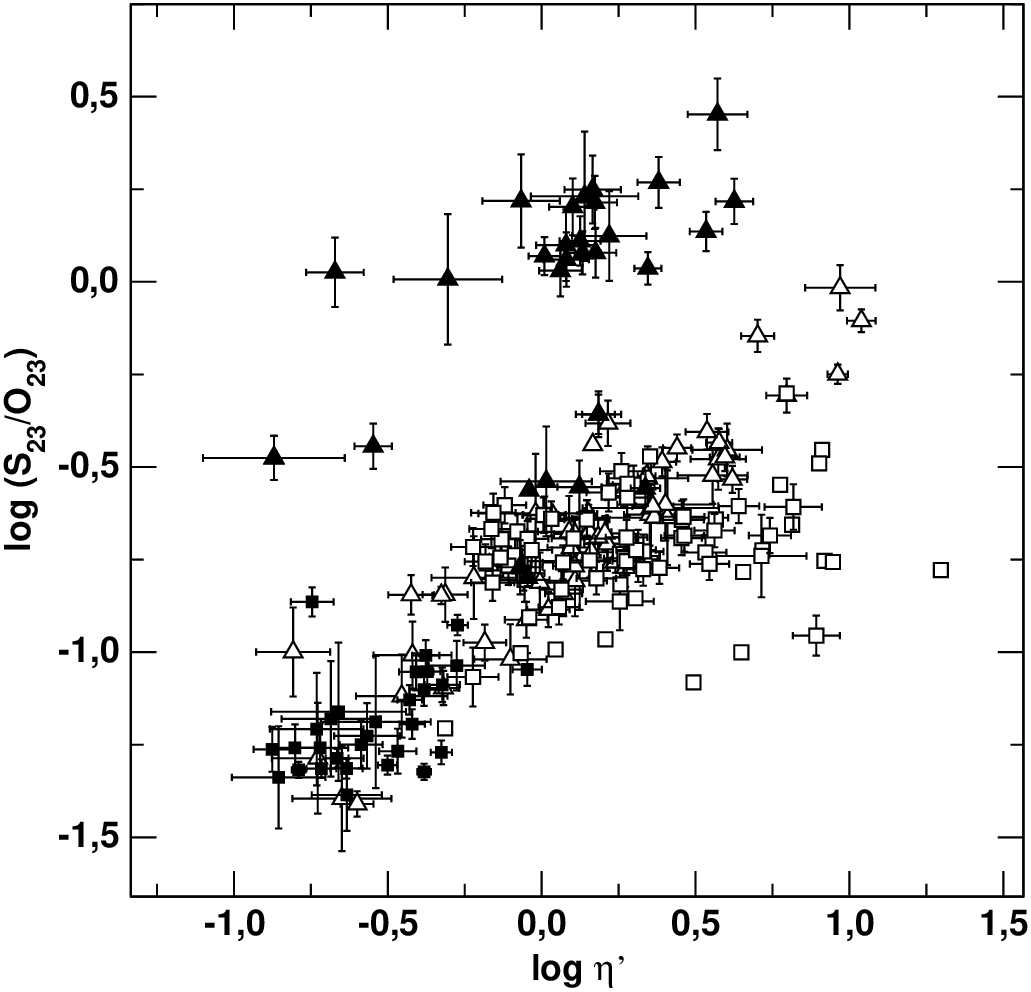}
  \end{center}
\caption{Relation between S23/O23 and the $\eta$' parameter. CNSFRS break the expected
relation between metallicity and effective equivalent temperature.}
\end{figure}

\section{Photo-ionisation models}
There are two main problems that we have tried to adress through the use of 
photo-ionisation models. Firstly, the searching for an explanation for the unexpected 
large effective temperatures found in CNSFRSs independently of their metallicites. 
Secondly, the fact that the ionising spectra corresponding to the temperatures in these objects and 
in HII galaxies is harder than the hardest available cluster theoretically predicted 
spectra. For this purpose we have used CLOUDY 96 (Ferland, 2002), with Starburst 99 
spectral energy distributions (atmospheres from Pauldrach et al. (2001) for O and B stars 
and Hillier \& Miller (1998) for WR). 

In the $\eta$' diagram it is possible to scale Teff. with CoStar model stellar 
atmospheres. These are represented as solid lines in figure 4. 
Both CNSFRs and HII galaxies show equivalent temperatures higher than 
50 kK, even for different excitation degrees and metallicites. Other types of objects 
lie between 30 and 40 kK.
Open squares represent clusters models with SB99 atmospheres at 1.0 Myr. The ionisation 
structure of the nebula is dominated by these very young bursts in objects showing high 
Teff.

\begin{figure}[t]

 \begin{center}
    \vspace{1.2cm}
    \includegraphics[width=8.5cm]{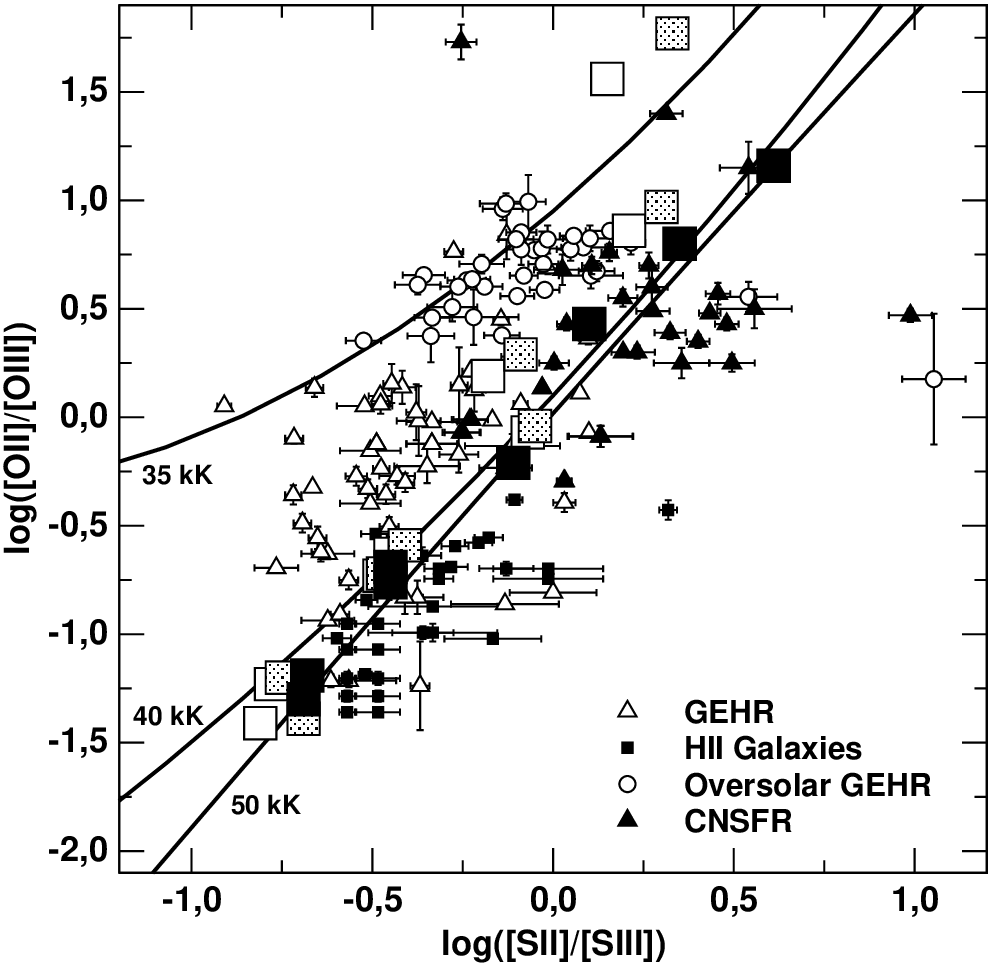}
  \end{center}
\caption{The $\eta$' diagram. The scale of equivalent effective temperatures, in 
solid lines, is compared with observations and cluster model atmospheres in photo-ionisation models,
represented by squares (see text for explanation).}
\vspace{-0.5cm}
\end{figure}

For each ionisation parameter, different metallicites have been chosen in order to 
simulate the gas conditions. These models have plane-parallel geometry and a
constant density of 100 cm$^{-3}$. They are summarized in table 1.

\begin{table}[h!]
\begin{center}
\begin{tabular}{cc}
\hline
{\bf log U} & {\bf Z} \\
\hline
-2.0 & 0.001, 0.004\\
-2.5 & 0.004, 0.008\\
-3.0 & 0.008, 0.020\\
-3.5 & 0.020, 0.040\\
\hline
\end{tabular}
\caption{Chosen metallicities for each ionisation parameters in the described models.}
\end{center}
\vspace{-0.7cm}
\end{table}

As SB99 models cannot reach the high effective temperatures deduced in CNSFRs and, 
to a lesser degree, in HII galaxies, we have tested to what extent the presence of all 
dust grain physics affects the equivalent temperature in the nebula. The models including 
dust grains are represented by dotted squares in figure 4. Although it is expected an increment of Teff 
at least at high metallicities, due to depletion factors and internal heating of the gas, 
there are only variations in the excitation degree, and in non-significant quantities. 

Finally, the geometry of the gas can also affect the inner ionisation equilibrium and the functional 
parameters may vary their physical meaning. In the models represented by filled squares in figure 4 we have 
put the gas near to the ionising source, resulting in a spherical geometry. At low values 
of the ionisation parameter and high metallicities, that is the range of CNSFRs, the 
effective equivalent temperature is increased. Nevertheless, at high excitation degrees 
and low metallicities, that is the range of HII galaxies, there are not significant 
variations. 

Therefore, although the mechanism of heating and ionisation of the gas in massive HII regions with very 
young and efficient processes of star formation in all ranges of metallicity is far 
from being well understood, the properties of the stellar ionising populations through 
direct observations in different bands and future more reallistic assumptions about 
the geometry of the gas will shed some light on this issue.


\begin{chapthebibliography}{1}
\bibitem{} \'Alvarez-\'Alvarez, M., D\'\i az, A.I., \& Castellanos, M. 2003, IAUS, 212, 537. 
\bibitem{} D\'\i az, A.I., Terlevich, E., V\'\i lchez, J.M., Pagel, B.E.J. \& Edmunds, M.G.  1991. MNRAS, 253, 245.
\bibitem{} D\'\i az, A.I. \& P\'erez-Montero, E. 2000. MNRAS, 312, 130.
\bibitem{} Ferland, G. 2002. HAZY, A brief introduction to CLOUDY. Univ. Kentucky.
\bibitem{} Gonz\'alez-Delgado, R., P\'erez, E., D\'\i az, A.I., Garc\'\i a-Vargas, M., Terlevich, E. \& V\'\i lchez, J.M. 1994, ApJ, 439, 604.
\bibitem{} Hillier, D.J. \& Miller, D.L. 1998, ApJ, 496, 407.
\bibitem{} Pastoriza, M.G., Dottori, H.A., Terlevich, E., Terlevich, R. \& D\'\i az, A.I. 1993, MNRAS, 260, 177.
\bibitem{} Pauldrach, A.W.A, Hoffmann, T.L. \& Lennon, M. 2001, A\&A, 375, 161.
\bibitem{} P\'erez-Montero, E. \& D\'\i az, A.I. 2004. in The Formation and Evolution of
Massive Young Star clusters. Eds. H.J.G.L.M. Lamers, L. J. Smith \& A. Nota. HASP conf. Series. v. 322, 213.
\bibitem{} P\'erez-Olea, D. PhD thesis. 1996. UAM.
\bibitem{} V\'\i lchez, J.M. \& Pagel, B.E.J. 1988. MNRAS, 231, 257.

\end{chapthebibliography}
\end{document}